\documentclass[12pt]{article}
\usepackage{graphicx}
\begin{document}
\def\be{\begin{equation}}
\def\ee{\end{equation}}
\thispagestyle{empty}
\begin{flushright}
CERN-TH/98-349\\ 
RIKEN-BNL Preprint\\
hep-ph/9811222
\end{flushright}
\vskip3cm

\centerline{\Large{\bf{The Glueball Filter in Central Production}}}
\vskip0.3cm
\centerline{\Large{\bf{and}}}
\vskip0.3cm
\centerline{\Large{\bf{Broken Scale Invariance}}}
\vskip2cm
\centerline{{\bf John Ellis$^{(a)}$} and {\bf Dmitri Kharzeev$^{(b)}$}}
\vskip0.5cm
\centerline{{\it $^{(a)}$ Theory Division, CERN, CH-1211 Geneva, Switzerland }}
\vskip0.3cm
%\centerline{{\it and}}
\centerline{{\it $^{(b)}$ RIKEN-BNL Research Center,}}
\centerline{{\it Brookhaven National 
Laboratory, Upton NY 11973, USA}}
\vskip1.5cm
\begin{abstract}
We propose a possible explanation of the kinematical dependence of the central 
production of the scalar glueball candidate observed recently by the
WA91 and WA102 
Collaborations, and discussed by Close and Kirk, in the context of the
broken scale invariance of QCD. The dependences of glueball
production on the transverse momenta
and azimuthal angles of the final-state protons
may be related to the structure of the trace anomaly
in QCD.
\end{abstract}

\vskip1cm
\begin{flushleft}
CERN-TH/98-349\\
RIKEN-BNL
\end{flushleft}

\newpage

Confinement and the non-Abelian structure of QCD imply the existence
of bound states 
of gluons. Clearly, finding and recognizing such glueball states is 
very important. One intriguing possibility is to identify the 
observed $f_0(1500)$ state with the lightest scalar glueball \cite{Rolf}. 
To verify the gluonic nature of this state, one has to check in particular 
if the mechanisms of its production are consistent with those expected 
for glueball states. This suggests in particular that one looks
for its production in gluon-rich environments.
It was suggested long time ago \cite{Robson} that the glueballs 
should be produced copiously in the central production process 
\be
pp \to p_f X p_s
\ee                            
This may be dominated by double-Pomeron exchange
when the final-state protons carry large fractions of the initial-state 
proton momenta in the centre-of-mass frame.
In fixed-target experiments, this requires the presence 
of fast ($p_f$) and slow ($p_s$) protons in the final state.  
\vskip0.3cm

Recently, a big step in the investigation of this process 
has been taken by the WA91 and WA102 Collaborations, which 
have reported remarkable 
kinematical dependences of central meson production~\cite{WA91,WA102}. 
Specifically, it was observed~\cite{WA102} that the 
the production of glueball candidates
depends strongly on the relative transverse
momenta of the final-state protons $p_f$ and $p_s$.
The variable suggested in~\cite{WA102,Close} was the 
difference between the transverse momenta $\vec{p\prime}_t$ and
$\vec{q\prime}_t$ of the final-state protons:
\be
dP_t = |\vec{p_t}\prime - \vec{q_t}\prime|.
\ee
The dependence of central meson production on this variable appears to be very 
non-trivial: namely, it was found that at small $dP_t$ the production 
of glueball candidates, in particular the $f_0(1500)$, was
significantly 
enhanced over the production of conventional $\bar{q}q$ mesons\footnote{
The cross section of conventional $\bar{q}q$ meson production 
by itself was found to be 
suppressed at small $dP_t$; for a theoretical interpretation, 
see Ref \cite{Frere}.}. 
It was proposed~\cite{Close} that this remarkable feature of central 
production could be related to the 
intrinsic structure of glueball states, and 
that the selection of events 
with small $dP_t$ could act effectively as a glueball filter. 
So far, no dynamical explanation of this important empirical observation 
has been suggested, so the challenge for theory is to understand
the  dynamics behind this glueball filter. 
\vskip0.3cm

In this Letter, we suggest a possible dynamical explanation of this 
empirical observation, based on the concept of broken scale invariance 
in QCD~\cite{cg,JE,JE1}. This framework requires the existence of a scalar
glueball,
which plays the role of the dilaton, 
saturating the matrix elements of the trace of the energy-momentum 
tensor of QCD $\Theta_{\mu}^{\mu}$. This operator includes an anomalous
piece containing gluon field strengths, and we propose that the
kinematic structure of the effective Pomeron-Pomeron-glueball vertex
may reflect that of the gluon-gluon piece in $\Theta_{\mu}^{\mu}$,
which is proportional to $F^{\mu\nu}F_{\mu\nu}$.
We demonstrate that this mechanism reproduces qualitatively the
observed dependence of the candidate scalar glueball production
on $dP_t$ and the relative angles of $ \vec{p_t}\prime ,
\vec{q_t}\prime$.

\vskip0.3cm

In central production at sufficiently high energies, mesons are
believed to be produced via Pomeron-Pomeron 
fusion, as shown in the top two diagrams of Fig.~1, although other
models are
possible, such as
that illustrated in the bottom diagram of Fig.~1. The Pomeron is known
to
couple to light
hadrons effectively 
as a vector 
particle~\cite{Wu,Land}. Moreover, it is also 
generally accepted that the Pomeron has 
a large gluon component, and this picture is supported by
analyses of diffraction at HERA. The strongest form of
this hypothesis is
the leading-gluon model, which postulates a hard 
distribution of gluons inside the Pomeron, as suggested by 
the H1 Collaboration~\cite{H1}. Within this model,
it is natural to describe meson production
in Pomeron-Pomeron collisions via the fusion 
of two leading gluons from the interacting Pomerons, as shown in the
central diagram of Fig.~1. 
Other interpretations of the H1 and ZEUS data are possible~\cite{BEKW},
but the success in other applications of the vector-dominance
model for Pomeron couplings~\cite{FEC} also motivates the suggestion that
the central production of 
scalar glueballs occurs via the coupling between the scalar dilaton field
and two vector fields. 
\vskip0.3cm
  
This coupling has been known since 1972, when the concept of 
the canonical trace anomaly was introduced~\cite{cg, JE, JE1}. 
This concept was later given 
legitimacy by QCD, which was explicitly shown to 
possess an anomalous term $\propto F^{\mu\nu}F_{\mu\nu}$ in the
energy-momentum tensor~\cite{scale}. If one assumes that
matrix elements of the trace of the energy-momentum tensor
are dominated by a scalar glueball field $\Theta$,
its resulting coupling to two vector fields would also have 
the form $\sim \Theta F^{\mu\nu}F_{\mu\nu}$. 
In an effective theory where the effective coupling of the Pomeron is 
that of a vector particle, $F^{\mu\nu}$ can be considered as an effective 
`Pomeron field'. In the leading-gluon model of the Pomeron structure, 
this is closely related to the gluon field strength $G_a^{\mu\nu}(x)$.

\vskip0.3cm

Based on the above arguments, we propose
the following form for the coupling responsible for scalar glueball
production in Pomeron-Pomeron collisions:
\be
\L \sim \Theta (x)G^{\mu\nu}(x)G_{\mu\nu}(x), 
\label{vert}
\ee 
In momentum space, this 
coupling leads to an amplitude that is proportional to the 
square of the scalar
product of the four-momenta of the colliding gluons $g_1$ and $g_2$:
\be
{\mathcal{M}} \sim (\epsilon^{\mu} g_1^{\nu} - \epsilon^{\nu} g_1^{\mu}) 
(\epsilon_{\mu} g_{2 {\nu}} - \epsilon_{\nu} g_{2 {\mu}})  
 \sim (g_1 . g_2).
\label{melement}
\ee 
whose implications for the WA91 and WA102 experiments we
now evaluate, assuming that the Pomeron-Pomeron-glueball vertex
has a similar structure.

\vskip0.5cm

Denoting the initial and final four-momenta of the colliding
protons by $p$,$q$ and $p\prime$, $q\prime$, respectively, 
and denoting their initial c.m.s. momentum by $P$, 
we can write
$$
p \simeq \left( P + {M^2 \over {2 P}},\ \vec{p_t} = 0,\ p_L = P \right),
$$
$$
q \simeq \left( P + {M^2 \over {2 P}},\ \vec{q_t} = 0,\ q_L = - P \right)
$$
and
$$
p\prime \simeq \left( x_1 P + {M^2 \over {2 x_1 P}},\ \vec{p_t}\prime,\  
p_L\prime = x_1 P \right),
$$
$$
q\prime \simeq \left(x_2 P + {M^2 \over {2 x_2 P}},\  \vec{q_t}\prime,\  
q_L\prime = - x_2 P \right).
$$
%\vskip0.3cm
%%%%%%%%%%%%%%%%%%%%%%%%%%%%%%%%5
\begin{figure}[h!]
\includegraphics[40mm,90mm][160mm,240mm]{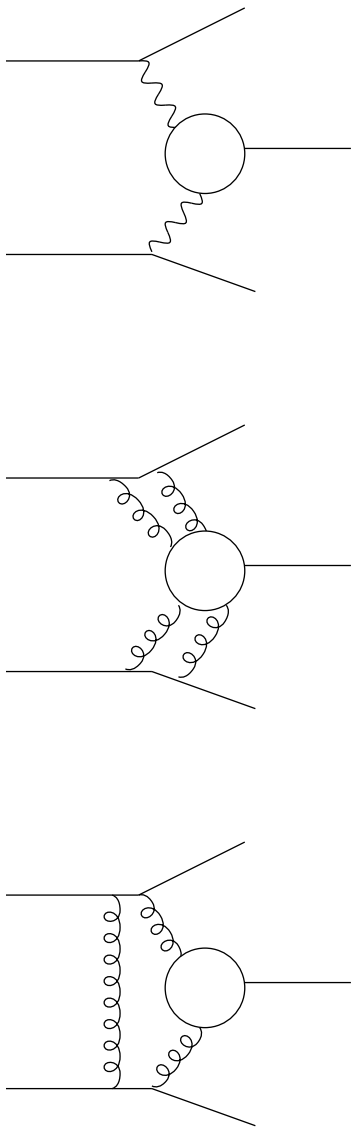}
\vskip0.4cm
\caption{Diagrams describing double-diffractive 
production. Top diagram: Pomeron-Pomeron fusion, with the zigzag
lines
denoting the Pomeron exchanges. Central diagram: as previously,
with two-gluon models for the Pomerons. Bottom diagram: another pattern of
gluon exchange that may lead to the same form of glueball production
vertex.}
\end{figure}
%%%%%%%%%%%%%%%%%%%%%%%%%%%%%%%%
\vskip0.3cm
\noindent
We now assume that the dependence of the production vertex on the
Pomeron momenta is proportional to that on the gluon momenta
in (\ref{melement}). This assumption would be trivial
in the leading-gluon model, since the
colliding gluons would carry essentially all of the transfered momentum,
with the other 
gluon(s) in the Pomerons
merely compensating the colour, which is assumed not to alter
the kinematic structure:
\be
g_1. g_2 \propto (p-p\prime). (q-q\prime) = 
 2 P^2 (1-x_1) (1-x_2) - p_t\prime. q_t\prime \  cos \phi, \label{momenta}
\ee 
where $\phi$ is the azimuthal angle between the directions of
the final-state protons. This proportionality assumption may also
hold in a more general approach to the vector-like couplings of
the Pomeron, and even in the different production mechanism
displayed in the bottom diagram of Fig.~1.
Making this proportionality
assumption, the central glueball production
rate contains a factor
\be
\hat{\sigma} \sim |{\mathcal{M}}|^2 \sim (g_1  g_2)^2,
\label{rate}
\ee
%This suggests that the production of the scalar glueball 
%should be most efficient when the two protons scatter in
%parallel or 
%antiparallel directions in the transverse plane, and with significant
%tranverse momenta, although
%these should not exceed the limit beyond which the
%validity of the Pomeron-Pomeron fusion picture breaks down.

\vskip0.3cm
In addition to the rate (\ref{rate}) for the glueball production
subprocess, one must also take into account the kinematics and
the appropriate Pomeron flux factors.
In particular, we must ensure the
mass-shell condition
\be
[(p - p\prime) + (q - q\prime)]^2 = m^2,
\ee
where $m$ is the glueball mass. In terms of our kinematical variables, 
this condition may be rewritten as
\be
\left( P (2-x_1-x_2) + {M^2 \over {2P}} \left(2- {1\over x_1} - {1\over x_2}
\right)
\right)^2 - (\vec{p_t\prime} + \vec{q_t\prime})^2 - (x_2 - x_1)^2 P^2 = m^2. 
\label{kin1}
\ee
When $x_1=x_2= x$ is close to unity, one has
\be
s (1-x)^2 - (\vec{p_t\prime} + \vec{q_t\prime})^2 \simeq m^2. \label{kin2}
\ee
This requirement must be combined with the kinematic dependence
of the production rate (\ref{rate}). Assuming that $x_1 = x_2 = x$,
and recalling that $t_{1,2}\simeq -
\vec{p_t}\prime^2, \vec{q_t}\prime^2$, we can rewrite (\ref{momenta}) as 
\be
g_1. g_2 \propto (p-p\prime). (q-q\prime) = 
{1 \over 2} \left(m^2 + 2\ |t| + 2\ |t|\ cos \phi \right) 
- |t|\ cos \phi.
\ee
The cross section (\ref{rate}) then takes the form
\be
\hat{\sigma} \sim (g_1.  g_2)^2 = \left[{m^2 \over 2} + |t| \right]^2.
\label{ratef}
\ee
Note that the fact that the dependence on the azimuthal angle $\phi$ has 
cancelled in (\ref{ratef}) is the consequence of the assumption 
$x_1=x_2$: in more general kinematic conditions, (\ref{ratef}) would 
depend on $\phi$. 
\vskip0.3cm
%%%%%%%%%%%%%
\begin{figure}[h]
%\begin{center}
{\Large{${d\sigma \over d\phi}$ \hskip65mm ${d\sigma \over d\phi}$}}

\includegraphics[width=140mm]{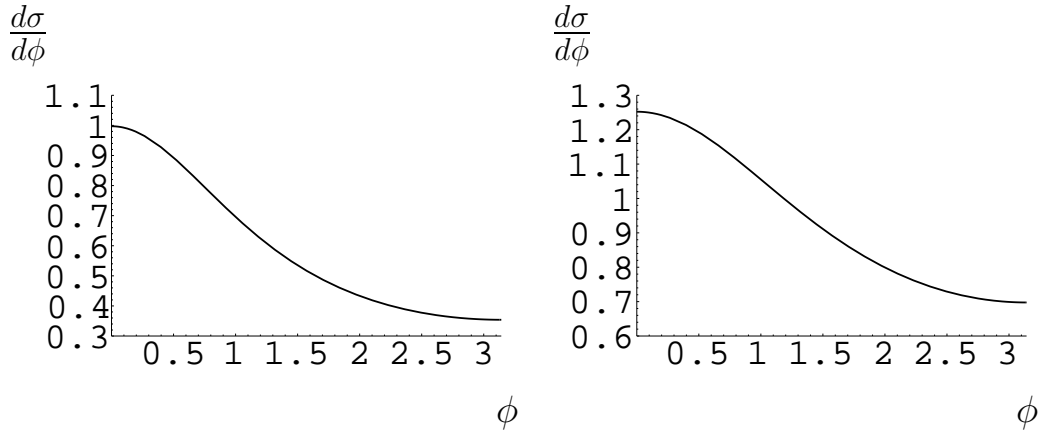}

\hskip65mm {\large{$\phi$}} \hskip65mm {\large{$\phi$}}
\caption{Distribution in the azimuthal angle $\phi$ (in radians) 
between the final-state
protons in double-diffractive production of the $f_0(980)$ (left panel) and 
$f_0(1500)$ (right panel) states, calculated for the center-of-mass
energy of the WA91 and WA102 experiments, and with $t = - 0.5$~GeV$^2$
for both final-state protons.}
%\label{prod}
\end{figure}
%%%%%%%%%%%%%%%%%
Assuming for simplicity that $x_1 = x_2 = x \equiv x_P - 1$, one can write
the cross section for double-diffractive glueball production in the 
following form:
\be
{\frac{d^2\sigma} {dt_1 dt_2 d\phi}} =  f_P(x_P,t_1)  f_P(x_P,t_2) 
{\frac{d^2\hat{\sigma}} {dt_1 dt_2 d\phi}}, \label{genf}
\ee 
where $\hat{\sigma}$ is defined in (\ref{rate}) and the Pomeron flux 
factors are given by \cite{DL}
\be
f_P(x_P,t) = {{9 \beta_0^2} \over {4 \pi^2}} \left[ F_1(t) \right]^2 
x_P^{1-2\alpha(t)},
\ee
with the Pomeron trajectory $\alpha(t)=1 + \epsilon + \alpha\prime t$, 
where $\epsilon \simeq 0.085$ and $\alpha\prime \simeq 0.25 ~ \rm{GeV}^2$, 
and $|F_1(t)|$ is the elastic form factor of the proton.
The value of $x_P$ is fixed by the kinematical constraint (\ref{kin2}):
\be
x_P = {\left( m^2 - t_1 - t_2 + 2\sqrt{t_1 t_2}\ cos \phi \right)^{1/2} \over
\sqrt{s}}.
\ee 
In the case when $t_1\simeq t_2 = t$, one can write down a simple formula 
for the distribution in the relative azimuthal angle $\phi$ :
\be
{d\sigma \over d\phi} \sim \left( s \over m^2 + 2 |t| (1 - cos \phi) \right)^
{1 + 2 \epsilon + 2 \alpha\prime t}\ \label{result}
\ee
The distribution (\ref{result}) for $f_0(980)$ and $f_0(1500)$
production in the WA91 and WA102 experiments when
$t=-0.5\ \rm{GeV}^2$ is shown in
Fig.~2, and
has an interesting feature --
it does favour the production 
of the glueball when the transverse momenta of the outcoming protons 
are parallel, in qualitative accord with the experimental observation 
\cite{WA91,WA102}. 
%We would expect some aspects of our calculation to be washed
%out in a more realistic model calculation which takes account of
%the internal structure of the Pomeron, deviations from the
%leading-gluon picture, etc.. 
We note that the
angular dependence we find is very different from that found
in~\cite{FEC} for other mesons. It would be
interesting to
study this azimuthal-angle 
dependence in more detail experimentally,
remembering that the detailed 
shape of the azimuthal dependence should depend on the kinematic 
selection, according to the expression (\ref{genf}). 
It would also be interesting to extend the studies of azimuthal-angle 
dependences in double diffractive processes to collider (RHIC or LHC) 
energies, 
where the dominance of the Pomeron exchange is better justified.      
    
{\bf Acknowledgements}

We thank Frank Close for comments on the earlier version of this paper.

\end{document}